\documentclass{WileyMSP-template}

\usepackage{siunitx}

\begin{document}

\pagestyle{fancy}
\rhead{\includegraphics[width=2.5cm]{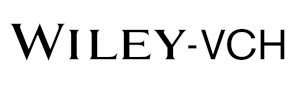}}

\title{High Resolution Temperature-Resolved Spectroscopy of the Nitrogen Vacancy $\mathrm{^{1}E}$ Singlet State Ionization Energy}

\maketitle


\author{Kristine V. Ung}
\author{Connor A. Roncaioli}
\author{Ronald L. Walsworth}
\author{Sean M. Blakley}


\dedication{}


\begin{affiliations}

Kristine V. Ung\\
Department of Physics, University of Maryland\\
College Park, Maryland, USA 20742\\
Joint Quantum Institute, University of Maryland\\
College Park, Maryland, USA 20742\\
Quantum Technology Center, University of Maryland\\
College Park, Maryland, USA 20742\\
kvung@umd.edu\\

Dr. Connor A. Roncaioli\\
DEVCOM Army Research Laboratory\\
2800 Powder Mill Rd, Adelphi, Maryland, USA 20783\\
connor.a.roncaioli.civ@army.mil\\

Dr. Ronald L. Walsworth\\
Department of Physics, University of Maryland\\
College Park, Maryland, USA 20742\\
Department of Electrical Engineering, University of Maryland\\
College Park, Maryland, USA 20742\\
Quantum Technology Center,
College Park, Maryland, USA 20742\\
walsworth@umd.edu\\

Dr. Sean M. Blakley\\
DEVCOM Army Research Laboratory\\
2800 Powder Mill Rd, Adelphi, Maryland, USA 20783\\
sean.m.blakley2.civ@army.mil\\

\end{affiliations}



\begin{abstract}

The negatively charged diamond nitrogen-vacancy ($\mathrm{{NV}^-}$) center plays a central role in many cutting edge quantum sensing applications; despite this, much is still unknown about the energy levels in this system. The ionization energy of the $\mathrm{^{1}E}$ singlet state in the $\mathrm{{NV}^-}$ has only recently been measured at between 2.25 eV and 2.33 eV. In this work, we further refine this energy by measuring the $\mathrm{^{1}E}$ energy as a function of laser wavelength and diamond temperature via magnetically mediated spin-selective photoluminescence (PL) quenching; this PL quenching indicating at what wavelength ionization induces population transfer from the $\mathrm{^{1}E}$ into the neutral $\mathrm{{NV}^0}$ charge configuration. Measurements are performed for excitation wavelengths between 450 nm and 470 nm and between 540 nm and 566 nm in increments of 2 nm, and for temperatures ranging from about 50 K to 150 K in 5 K increments. We determine the $\mathrm{^{1}E}$ ionization energy to be between 2.29 and 2.33 eV, which provides about a two-fold reduction in uncertainty of this quantity.  Distribution level: A. Approved for public release; distribution unlimited.

\end{abstract}

\medskip
\section{Introduction}

\begin{figure}
\centering
\includegraphics[width=0.5\linewidth]{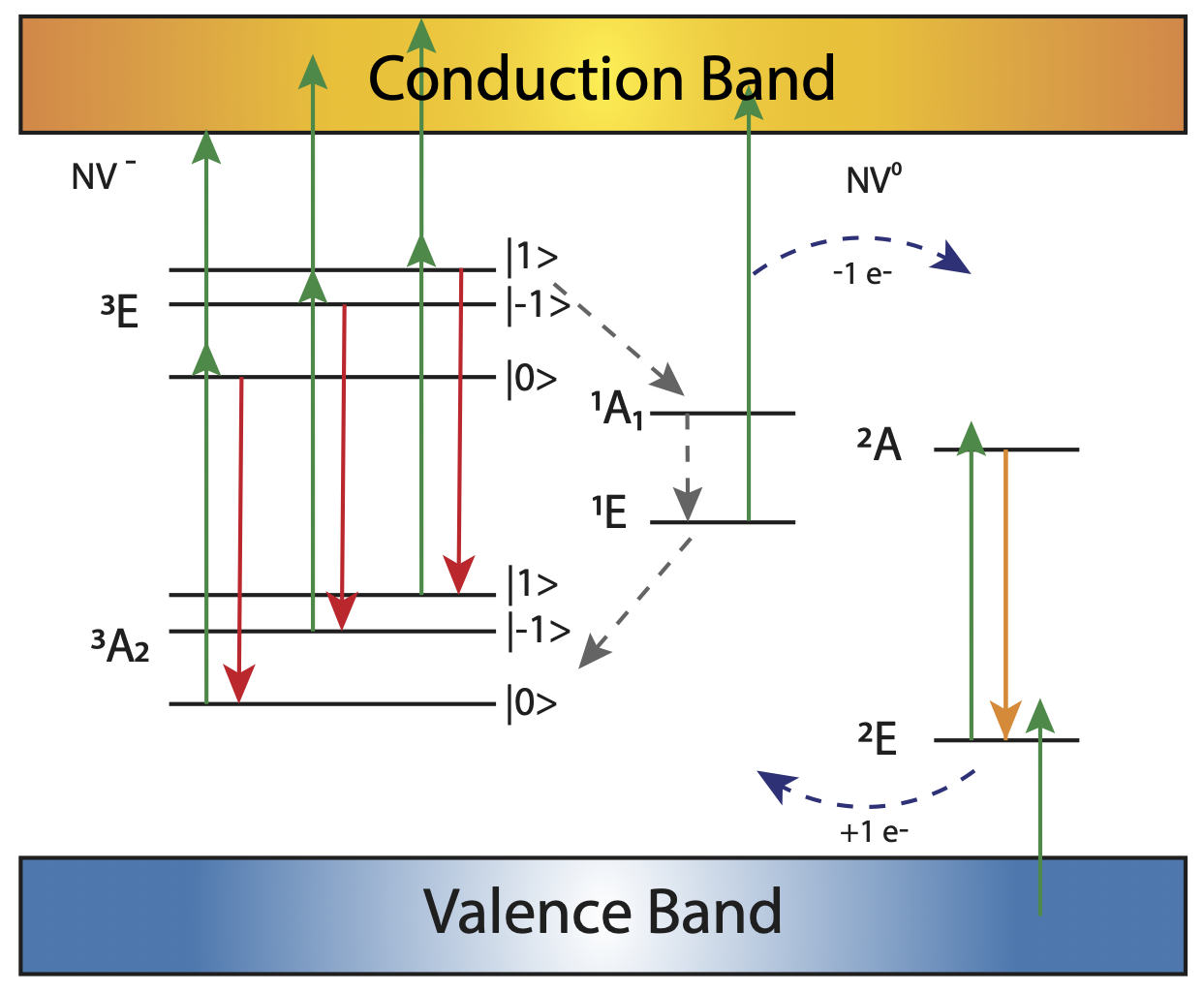}
  \caption{Energy level diagram of the nitrogen--vacancy system in the negative ($\mathrm{{NV}^-}$) and neutral ($\mathrm{{NV}^0}$) charge configurations. Solid straight arrows indicate transitions mediated by one photon, whereas dashed straight arrows indicate transitions mediated primarily by phonons.  Curved dashed arrows indicate transitions between charge configurations.}
  \label{fig:energy-diagram}
\end{figure}

The nitrogen vacancy (NV) center in diamond has proven to be a highly effective quantum sensor due to its good thermal properties,\cite{Wojciechowski_PrecisionTemp_2018,yang_diamond_2019,blakley_stimulated_2016} stable fluorescence,\cite{fedotov_background-free_2019} and high spin contrast,\cite{barry_sensitivity_2020} making it a sensitive and adaptable measurement tool for a variety of applications.\cite{schirhagl_nitrogen-vacancy_2014,doherty_towards_2016} Understanding the energy level structure of the NV center has been critical for quantum sensing with the NV center, enabling the development of measurement techniques like spin-to-charge conversion and photoelectric detection of magnetic resonance (PDMR) that possess a higher contrast and signal count rates relative to typical fluorescence sensing.\cite{barry_sensitivity_2020,bourgeois_photoelectric_2015,bourgeois_enhanced_2017,bourgeois_photoelectric_2020,bourgeois_photoelectric_2022} The $\mathrm{^{1}E}$ singlet state in the negative $\mathrm{{NV}}$ charge configuration ($\mathrm{{NV}^-}$) is of key importance to optically detected magnetic resonance (ODMR) quantum sensing with the $\mathrm{{NV}}$ center, and yet it remains the least well understood of the known energy levels.  Previous work on this subject has shown, via direct photoionization, that the ionization energy of the $\mathrm{^{1}E}$ is between 2.25 eV and 2.33 eV.\cite{Blakley_photoionization2024} These measurements were performed with a combination of gas, dye, and solid state lasers, and possessed an average wavelength resolution of approximately 13 nm and an average temperature resolution of approximately 50 K.  This course resolution left a large area of uncertainty in the measured value of the $\mathrm{^{1}E}$ ionization energy. In this manuscript, we report a two-fold reduction in the uncertainty in the $\mathrm{^{1}E}$ ionization energy by measuring PL spectra over two wavelength ranges, 450-470 nm and 540-566 nm, with 2 nm wavelength increments. Measurements are performed at two laser powers of 0.1 mW and 1.0 mW, at 5 K temperature steps from 50 K to 150 K. 

\section{Experimental Section}

 The $\mathrm{{NV}^-}$ charge configuration has four electronic spin states which include the triplet ground state ($\mathrm{^{3}A_2}$) and the triplet excited state ($\mathrm{^{3}E}$) both of which have a S=1 spin \cite{doherty_nitrogen-vacancy_2013,gali_recent_2023,jelezko_single_2006}, as well as a singlet (S=0) excited state ($\mathrm{^{1}A_1}$) and the metastable $\mathrm{^{1}E}$ singlet state. When illuminating the system with wavelengths between 470 and 637 nm, an NV absorbs a photon and transitions from the $\mathrm{^{3}A_2}$ to the $\mathrm{^{3}E}$ while conserving electronic spin. The system can either relax back to the $\mathrm{^{3}A_2}$ via photon emission (PL) or into the $\mathrm{^{1}A_1}$ via the upper inter-system crossing (ISC) and then into the $\mathrm{^{1}E}$.  The transition into the $\mathrm{^{1}E}$ via the upper ISC and the $\mathrm{^{1}A_1}$ is a non radiative process in the visible wavelength band. For the $\mathrm{{m}_z = \pm 1}$ spin sublevels of the $\mathrm{^{3}E}$, relaxation to $\mathrm{^{1}E}$ via the upper ISC is more likely than relaxation to $\mathrm{^{1}E}$ from the $\mathrm{^{3}E}$ $\mathrm{{m}_z = 0}$ spin sublevel. The decay rate back to $\mathrm{^{3}A_2}$ from the $\mathrm{^{1}E}$ via the lower ISC is relatively slow, giving the $\mathrm{^{1}E}$ a lifetime that is an order of magnitude larger than any other NV state besides the $\mathrm{^{3}A_2}$.\cite{acosta_optical_2010} Under suitable optical illumination, these processes lead to the preparation of the $\mathrm{{NV}^-}$ electronic spin state into $\mathrm{{m}_z = 0}$ state, and also allow the $\mathrm{{NV}^-}$ spin state to be readout via PL (Fig. \ref{fig:energy-diagram}). While the photophysics of most $\mathrm{{NV}^-}$ states has been well established for a number of years \cite{aslam_photo-induced_2013,beha_optimum_2012} $\mathrm{^{1}E}$ ionization has been the subject of debate in the literature\cite{goldman_erratum_2017,goldman_phonon-induced_2015,goldman_state-selective_2015,toyli_measurement_2012,manson_nv_2018}, with the $\mathrm{^{1}E}$ ionization energy only recently determined.\cite{Blakley_photoionization2024,wolf_nitrogen-vacancy_2022,razinkovas_photoionization_2021,aude_craik_microwave-assisted_2020,hacquebard_charge-state_2018,bockstedte_ab_2018,thiering_theory_2018} 

To measure the $\mathrm{^{1}E}$ ionization energy, a wavelength-resolved direct photoionization method was developed.\cite{Blakley_photoionization2024} This method involves applying a bias magnetic field of at least 100 mT along the diamond $\mathrm{<100>}$ axis so that the $\mathrm{m_s=0}$ and $\mathrm{m_s=-1}$ spin sublevels\cite{ernst_temperature_2023,ernst2023modeling,chakraborty_magnetic-field-assisted_2022} in the $\mathrm{^3E}$ triplet excited state are equally mixed for $\mathrm{{NV}^-}$ centers along all 4 diamond crystallographic axes.  The choice of this field magnitude and direction ensures an increase in the population of the $\mathrm{^{1}E}$ via its ISC with $\mathrm{^3E}$, resulting in a measurable increase in $\mathrm{{NV}^0}$ PL due to $\mathrm{^{1}E}$ photoionization.

\subsection{Measurement Procedure}

\begin{figure}
\centering
\includegraphics[width=\linewidth]{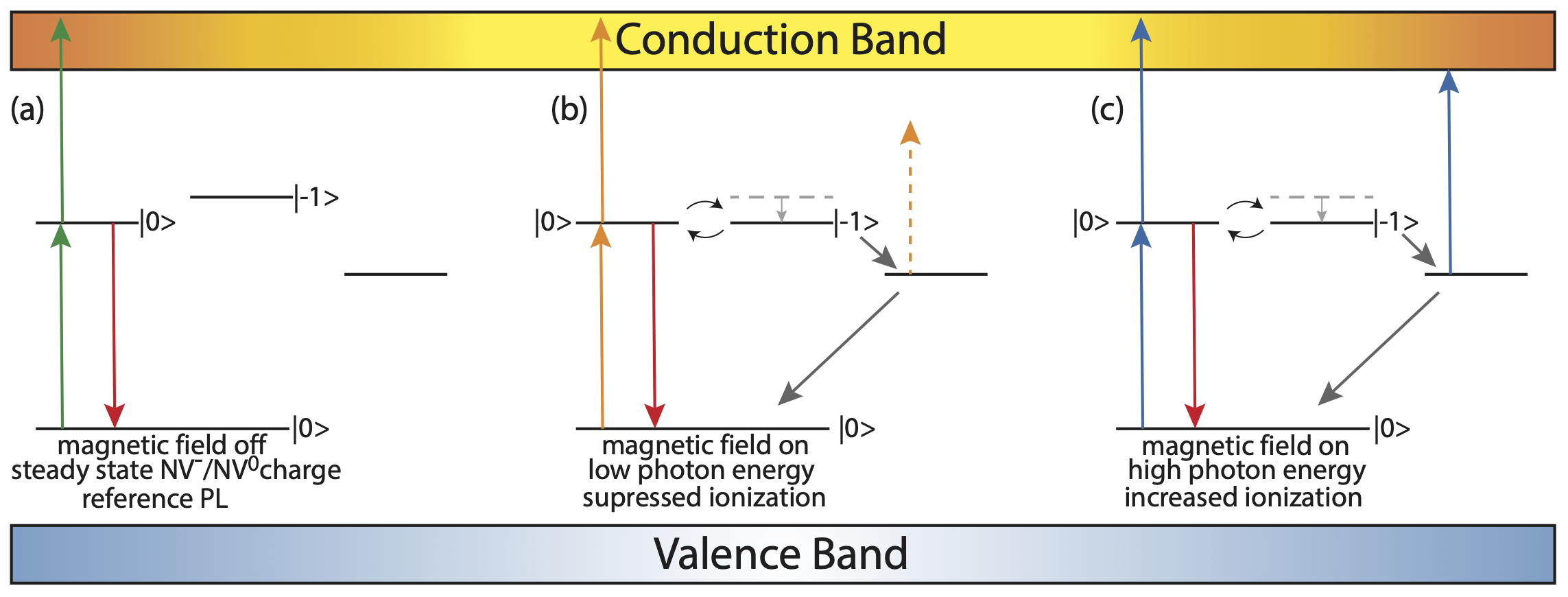}
  \caption{Ionization measurement sequence for spectroscopic analysis of $\mathrm{^{1}E}$ singlet. (a) Field-off reference PL measurement compared to field-on signal measurement (b, c). (b) If photon energy less than $\mathrm{^{1}E}$ ionization threshold, $\mathrm{^{1}E}$ protects $\mathrm{^{3}E}$ from Auger ionization; decreased $\mathrm{{NV}^0}$ PL output relative to reference. (c) If photon energy greater than $\mathrm{^{1}E}$ ionization threshold, $\mathrm{^{1}E}$ ionizes in addition to $\mathrm{^{3}E}$, increased $\mathrm{{NV}^0}$ PL output relative to reference.}
  \label{fig:ionize_process}
\end{figure}

The present work employs an expanded version of the method used in Ref. \cite{Blakley_photoionization2024}. The procedure for measuring the $\mathrm{^{1}E}$ energy proceeds in two steps.  The first step involves recording a reference PL spectrum from the $\mathrm{{NV}^-}$ centers with no bias magnetic field, and optical illumination from one candidate ionization wavelength (Fig. \ref{fig:ionize_process} (a)). This PL spectrum defines the reference measurement, creating a steady-state charge configuration ratio between $\mathrm{{NV}^-}$ and $\mathrm{{NV}^0}$ at the applied ionization wavelength for a system with negligible $\mathrm{^{1}E}$ population.  Due to the lack of population in $\mathrm{^{1}E}$, the only photoionization pathways available to the $\mathrm{NV}$ system for this combination of experimental parameters are either: direct single photoionization from $\mathrm{^{3}A_2}$ triplet ground state to the conduction band\cite{doherty_nitrogen-vacancy_2013,aslam_photo-induced_2013} for sufficiently large ionization energies, or two-photon Auger ionization proceeding from the $\mathrm{^{3}A_2}$ to the conduction band via the $\mathrm{^3E}$ triplet for lower ionization energies.\cite{aslam_photo-induced_2013} The second step involves recording a PL spectrum with a bias magnetic field of 100 mT for the same ionization wavelength as the reference measurement (Fig. \ref{fig:ionize_process} (b), (c)). The application of bias magnetic field populates the $\mathrm{^{1}E}$ state via spin-state mixing\cite{Blakley_photoionization2024}, opening an additional two-photon ionization pathway for the $\mathrm{NV}$ system, first by proceeding from $\mathrm{^{3}A_2}$ to $\mathrm{^{3}E}$ via one photon absorption; and then decaying to $\mathrm{^{1}E}$ via its ISC with $\mathrm{^{3}E}$. Once in $\mathrm{^{1}E}$, an $\mathrm{{NV}^-}$ can either ionize via absorption of a second photon of sufficient energy, or remain in  $\mathrm{^{1}E}$ (and later decay to $\mathrm{^{3}A_2}$) if the second photon energy is too small to trigger ionization.

If the photon energy (i.e., illumination wavelength) is insufficient for $\mathrm{{NV}^-}$ ionization from the $\mathrm{^{1}E}$ state, then $\mathrm{^{1}E}$ occupancy protects the $\mathrm{{NV}^-}$ from two-photon Auger ionization by reducing the steady-state population in the $\mathrm{^{3}E}$ state during optical illumination (Fig. \ref{fig:ionize_process} (b)) in a process typically referred to as ``$\mathrm{^{1}E}$ shelving". This shelving mechanism shifts the charge configuration ratio in favor of $\mathrm{{NV}^-}$, reducing the contribution of $\mathrm{{NV}^0}$ to the measured PL spectrum in comparison to the reference PL spectrum. If the photon energy is sufficient to ionize the $\mathrm{^{1}E}$, this results a further increase in the $\mathrm{{NV}^0}$ population, increasing its contribution to the measured PL spectrum when compared to the reference PL spectrum measured in the first step (Fig. \ref{fig:ionize_process} (c)). 

\subsection{Experimental Apparatus}

A Hubner C-Wave laser is used to induce $\mathrm{^{1}E}$ ionization of an ensemble of $\mathrm{{NV}^-}$ centers in the diamond sample. The laser consists of an oven-controlled, periodically polled lithium niobate (PPLN) nonlinear crystal in a bow-tie cavity with an in-line etalon. This cavity is pumped by the second harmonic output of a 5W diode-pumped frequency doubled Nd:YAG laser, forming an optically parametric oscillator (OPO) that is continuously tunable within two bands in the infrared (IR): from 900 nm to 1040 nm and from 1080 nm to 1300 nm. Wavelength tuning is accomplished by pumping the PPLN crystal at the polling period index associated with the desired output band, adjusting the PPLN crystal oven temperature to tune the output within the band selected by the polling period, and filtering the resulting output wavelengths with the in-line etalon.

Once locked to the desired IR wavelength, the OPO output light enters a second bowtie cavity with another in-line etalon where it is frequency doubled by an additional oven-controlled second-harmonic generation (SHG) PPLN crystal using the same tuning method as for the OPO. The combined optical output of the OPO and SHG system is thereby continuously tunable from 450 nm to 520 nm and from 540 nm to 650 nm in the visible wavelength band. The tuned SHG output is coupled to a polarization maintaining (PM) fiber as it exits the C-Wave laser module.

After tuning the laser output to the desired wavelength, the laser power is stabilized by a Glan-Taylor polarizer mounted inside a stepper-motorized rotation stage at the output of the PM fiber. The resulting tuned and stabilized laser power is then sampled by a 10\% pick-off beam splitter and coupled to a diode power meter for monitoring. The laser power stabilizer unit is controlled by a software module that rotates the polarizer to tune the power output to the user specified setpoint as read by the power meter, and then continuously adjusts the power once every 25 ms to keep the laser power within $\pm 0.25\%$ of the power setpoint value.

The wavelength and power tuned laser output then proceeds through a pair of steering mirrors and is reflected off a 570 nm long pass filter into a 10x long working distance optical objective focused through the input viewport at the top of a Montana Instruments cryostat and onto the $\mathrm{{NV}^-}$ diamond sample. The diamond is mounted along its $\mathrm{<100>}$ axis between the poles of a controllable uniaxial electromagnet (max field of $\pm$450 mT) and on top of a 3-axis Attocube piezo stack that moves the diamond into the focal point of the objective. The magnetic field magnitude and cryostat temperature are controlled by a software module that tunes both the field and temperature to user specified set points.  The 3 mm x 3 mm x 0.5 mm quantum grade $\mathrm{{NV}^-}$ diamond sample (CVD grown by Element Six) is isotopically enriched to 99.99\%.  $\mathrm{^{12}C}$ content, with [N]$\approx$ 16 ppm and [NV]$\approx$ 4 ppm, homogeneously distributed throughout the entire sample volume. The diamond plate is polished sp that the top face is along the <100> direction. The PL signal from the diamond is collected by the optical objective and directed through the 570 nm long pass filter into the fiber input of a Ocean Optics QE Pro low-noise wide-band spectrometer with a bandwidth between 350 nm and 1140 nm. PL spectra are collected at an exposure time of 100 ms and averaged 10 times before being saved by a software module for further analysis.

Signal and reference PL spectral pairs are collected for every combination of laser powers of 100 $\mathrm{\mu W}$ and 1 mW and wavelengths from 450 nm to 470 nm and from 540 nm to 566 nm in 2 nm increments. Every combination of laser power and wavelength is also measured as a function of diamond temperature at 5 K increments between 50 K and 150 K, for a total of 1050 PL spectral pairs across 21 diamond temperatures, 25 laser wavelengths, and 2 laser powers.

\begin{figure}
\centering
\includegraphics[width=0.5\linewidth]{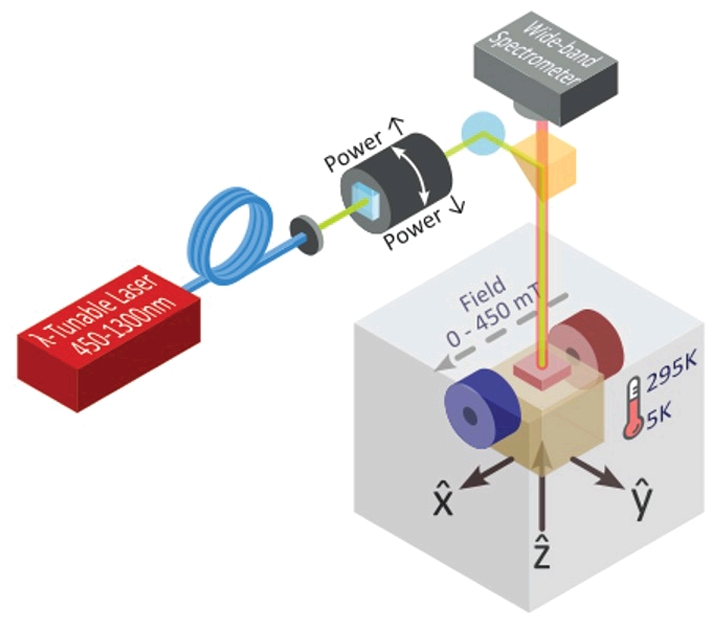}
  \caption{Experimental apparatus for measuring ionization threshold of $\mathrm{^{1}E}$ state, consisting of a wavelength-tunable continuous wave (CW) laser, a mechanical laser power tuner, a temperature controlled cryostat with 3-axis translation stage and optical access, a tunable uniaxial magnetic field, a $\mathrm{NV}$ diamond, a wide-band low-noise optical spectrometer, and associated optical elements.}
  \label{fig:apparatus}
\end{figure}


\section{Results \& Discussion}

\begin{figure}[ht!]
\centering
\includegraphics[width=0.75\linewidth]{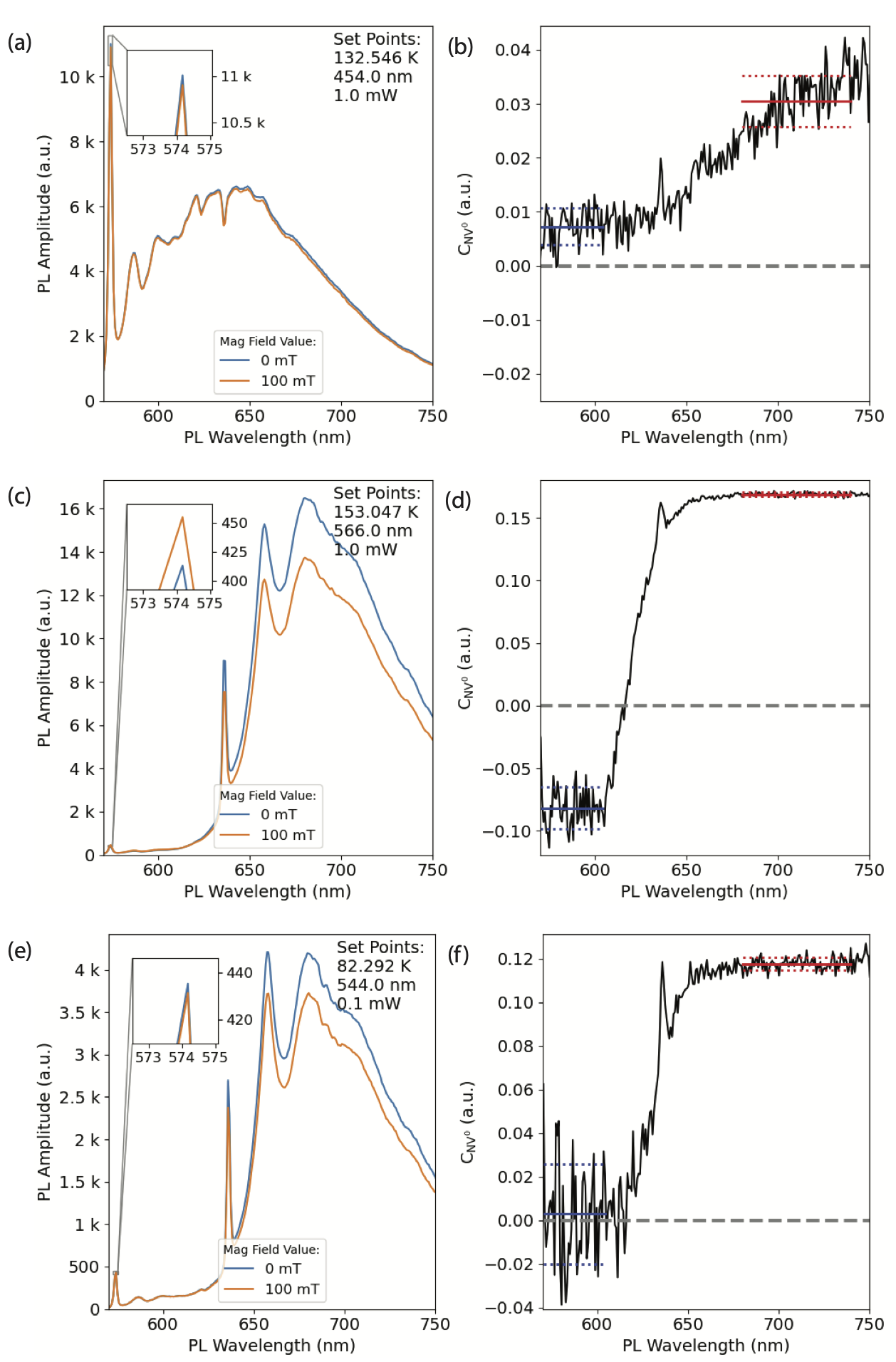}
  \caption{Representative measured PL spectral pairs (bias field on and off) in (a), (c), and (e) for indicated experimental conditions, and their calculated $C_{{NV}^0}$ values (b), (d), and (f), demonstrating $C_{{NV}^0} > 0$, $C_{{NV}^0} < 0$, and $C_{{NV}^0} \approx 0$, respectively. Solid (dashed) blue and red lines correspond to mean value (1 standard deviation) for $C_{{NV}^0}$ and ${{NV}^-}$ PL contrast respectively. For $C_{{NV}^0} > 0$, $C_{{NV}^0} < 0$, and $C_{{NV}^0} \approx 0$, the associated spectral pairs demonstrate decreased, increased, and mostly unchanged PL in the $\mathrm{{NV}^0}$ wavelength bands; this behavior can be seen in the inset figures in (a), (c), and (e). All possess the expected reduced PL in the $\mathrm{{NV}^-}$ band.}
  \label{fig:spectral_ratios}
\end{figure}

The $\mathrm{{NV}^0}$ PL amplitude is directly proportional to the relative $\mathrm{{NV}^0}$ population,\cite{Blakley_photoionization2024} and therefore serves as the primary experimental metric of changes in NV charge-state ratio in response to different experimental settings. In the wavelength region corresponding to significant $\mathrm{{NV}^0}$ PL emission, the ratio of the PL signal and reference spectra will become greater than one when the PL contribution from the $\mathrm{{NV}^0}$ is increased in the signal spectrum relative to the $\mathrm{{NV}^0}$ PL in the reference spectrum. Because the only difference in experimental settings between the signal spectrum and the reference spectrum is the presence of a magnetic field in the signal spectrum, a spectral ratio of greater than one in the $\mathrm{{NV}^0}$ band implies increased population in the $\mathrm{{NV}^0}$ charge configuration when the field is on, indicating that ionization of the $\mathrm{^{1}E}$ is occurring. Conversely, when this ratio is less than one, ionization is not occurring due to $\mathrm{^{1}E}$ shelving.

\begin{equation}
  \centering
  C_{{NV}^0}={\overline{I_{on}}/\overline{I_{off}}}-1
  \label{eq:ion_contrast}
\end{equation}

\begin{figure}
\centering
\includegraphics[width=\linewidth]{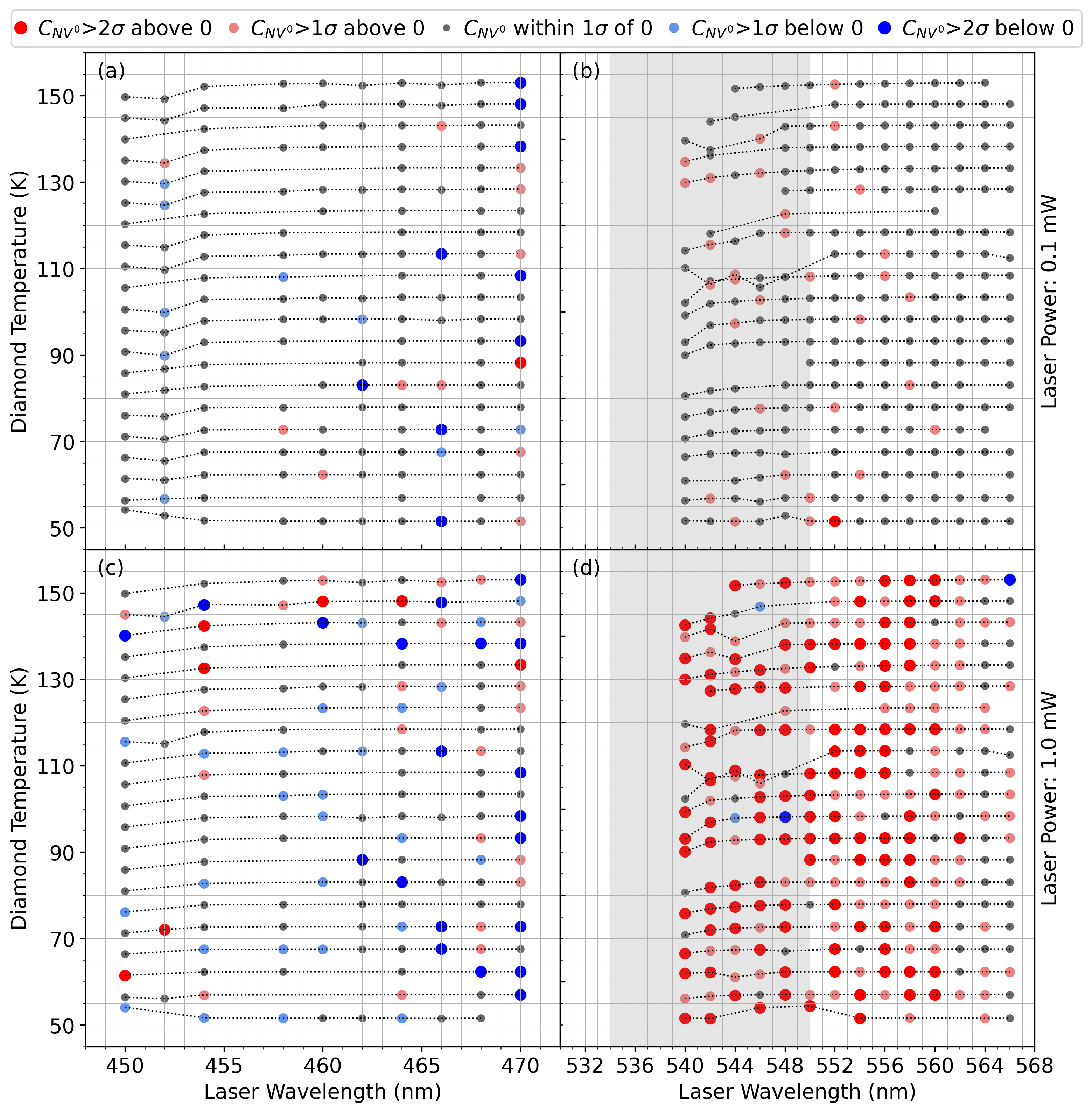}
  \caption{$C_{{NV}^0}$ as a function of temperature and wavelength for laser power (a), (b) 100 $\mu W$ and 1 mW (c), (d) over wavelength bands from 450 nm to 470 nm (a), (c), and 540 nm to 560 nm (b), (d).  Red circles indicate strong (large dark colored circles) or weak (small light colored circles) $\mathrm{^{1}E}$ shelving where $C_{{NV}^0}$ is greater than two standard deviations or greater than one but less than two standard deviations above zero respectively.  Small grey circles indicate that $C_{{NV}^0}$ is less than one standard deviation above zero, corresponding to indeterminate shelving or ionization behavior. Blue circles indicate strong (large dark colored circles) or weak (small light colored circles) $\mathrm{^{1}E}$ ionization where $C_{{NV}^0}$ is greater than two standard deviations or greater than one but less than two standard deviations below zero respectively. Dotted grey horizontal traces indicate measurements taken within a given cryostat temperature setpoint.  Grey shaded region in (b), (d) indicates region of uncertainty for $\mathrm{^{1}E}$ ionization energy determined in Ref. \cite{Blakley_photoionization2024}}
  \label{fig:contrast_plot}
\end{figure}

We define $C_{{NV}^0}$ in Eq. \ref{eq:ion_contrast} as the normalized NV ionization contrast for a given combination of experimental settings, where $I_{on} (I_{off})$ is the arithmetic mean of the relative measured $\mathrm{{NV}^0}$ PL for the bias field on (field off) setting over the wavelength band between 575 nm and 605 nm; negative values of $C_{{NV}^0}$ indicate ionization suppression due to $\mathrm{^{1}E}$ protecting the $\mathrm{{NV}^-}$ charge configuration from Auger ionization (i.e., $\mathrm{^{1}E}$ shelving), and positive values of $C_{{NV}^0}$ indicate $\mathrm{^{1}E}$ ionization. $C_{{NV}^0}$ is determined from measured of signal and reference PL spectral pairs for every combination of diamond temperature, laser wavelength, and laser power. Representative measured PL spectral pairs (Fig. \ref{fig:spectral_ratios} (a),(c), and (e)) and their associated $C_{{NV}^0}$ values (Fig. \ref{fig:spectral_ratios} (b),(d), and (f)) demonstrate $C_{{NV}^0}$ greater than, less than, and approximately equal to zero, respectively. Values of $C_{{NV}^0}$ greater than or less than zero are associated with reduced overall ionization due to $\mathrm{^{1}E}$ shelving or increased $\mathrm{^{1}E}$ ionization; and values of $C_{{NV}^0}$ within one standard deviation of zero are considered inconclusive $\mathrm{^{1}E}$ ionization behavior. (Note that this approach to PL data analysis is more efficient than that used in Ref. \cite{Blakley_photoionization2024}, which required Bayesian analysis to separate the $\mathrm{{NV}^{0}}$ PL spectrum from the total measured PL spectrum for calculating $C_{{NV}^0}$)  $C_{{NV}^0}$ values determined as in Fig. 4 are plotted as a function of temperature over wavelength bands from 450 nm to 470 nm and 540 nm to 566 nm (Figs. \ref{fig:contrast_plot} (a), (c), and (b), (d), respectively); and for laser powers of 100 $\mathrm{\mu W}$ and 1 mW (Figs. \ref{fig:contrast_plot} (a), (b), and (c), (d), respectively).  Values of $C_{{NV}^0}$ more than two standard deviations greater or less than zero are interpreted as temperature and wavelength regions corresponding to strong $\mathrm{^{1}E}$ shelving (Fig. \ref{fig:contrast_plot} larger dark red circles) or ionization (Fig. \ref{fig:contrast_plot} larger dark blue circles), respectively; whereas values of $C_{{NV}^0}$ that are between one and two standard deviations greater or less than zero represent regions of weaker $\mathrm{^{1}E}$ shelving (Fig. \ref{fig:contrast_plot} smaller pale red circles) or ionization (Fig. \ref{fig:contrast_plot} smaller pale blue circles), respectively. Regions where $C_{{NV}^0}$ is within one standard deviation greater or less than zero are considered to have inconclusive ionization or shelving behavior (Fig. \ref{fig:contrast_plot} tiny grey circles). For laser powers of 100 $\mathrm{\mu W}$ (Figs. \ref{fig:contrast_plot} (a), (b)), a few regions of strong $\mathrm{^{1}E}$ ionization occur at the edge of the triplet ionization threshold near 470 nm (2.63 eV) for temperatures above 100 K, but there is no indication of $\mathrm{^{1}E}$ ionization in the range between 540 nm (2.29 eV) and 566 nm (2.19 eV). At laser powers of 1.0 mW, there is further evidence of strong $\mathrm{^{1}E}$ ionization near the edge of the triplet ionization threshold at most temperatures between 50 K and 150 K (Fig. \ref{fig:contrast_plot} (c)).  For wavelengths between 540 nm and 566 nm, $\mathrm{^{1}E}$ shelving dominates at all temperatures (Fig. \ref{fig:contrast_plot} (d)), indicating almost no ionization of the $\mathrm{^{1}E}$ for wavelengths longer than 540 nm in this temperature range, with the notable exception for measurements for laser wavelengths of 548 nm and 566 nm at temperatures of 95 K and 150 K respectively. The physical phenomena that gives rise to these two points is unknown and warrants further study.

This result considerably narrows the region of uncertainty for the onset of $\mathrm{^{1}E}$ ionization from between 532 nm (2.33 eV) and 550 nm (2.25 eV) demonstrated in previous work\cite{Blakley_photoionization2024} to 532 nm (2.33 eV) and 540 nm (2.29 eV) for diamond temperatures between 50 K and 150 K — a two-fold reduction in uncertainty compared to the result in Ref. \cite{Blakley_photoionization2024} (Fig \ref{fig:contrast_plot} (b), (d) shaded grey region).

\section{Conclusion} 

By measuring the ratio between NV PL spectra measured with and without a 100 mT bias magnetic field and for many of temperatures (50 to 150 K), laser excitation wavelengths (450 to 470 nm and 540 to 566 nm), and laser powers (0.1 and 1 mW), we find $\mathrm{^{1}E}$ ionization energy to be between 2.29 eV and 2.33 eV, representing a two-fold reduction in the uncertainty in this quantity.  This result further resolves the range of energies needed for direct spin-to-charge conversion from the $\mathrm{^{1}E}$ singlet, which is a well-known tool for improving the performance of quantum sensors that rely on optically detected or PDMR measurement techniques.



\medskip

%

\bibliographystyle{MSP}
\bibliography{main}








\end{document}